\documentclass{article}

 \usepackage[dblblindworkshop,final]{neurips_2025}

\workshoptitle{Differentiable Systems and Scientific Machine Learning (EurIPS)}

\usepackage[utf8]{inputenc} 
\usepackage[T1]{fontenc}    
\usepackage{hyperref}       
\usepackage{url}            
\usepackage{booktabs}       
\usepackage{amsfonts}       
\usepackage{nicefrac}       
\usepackage{microtype}      
\usepackage{xcolor}         
\usepackage{graphicx}
\usepackage{amsmath}
\usepackage{svg}
\usepackage{soul}
\usepackage{comment}

\title{A differentiable model of supply-chain shocks}

\author{%
  Saad Hamid\textsuperscript{1}\thanks{Authors listed alphabetically.} \\
  \texttt{saad.hamid@aioilab-oxford.eu}
  \And
  José Moran\textsuperscript{2,3} \\
  \texttt{jose.moran@macrocosm.group}
  \And
  Luca Mungo\textsuperscript{2,3} \\
  \texttt{luca.mungo@macrocosm.group}
  \And
  Arnau Quera\mbox{-}Bofarull\textsuperscript{2} \\
  \texttt{arnau@macrocosm.group}
  \And
  Sebastian Towers\textsuperscript{2,4,5} \\
  \texttt{sebastian@macrocosm.group}
  \\
  \\
  \textsuperscript{1}\,Aioi R\&D Lab
  \quad \textsuperscript{2}\,Macrocosm\\
  \textsuperscript{3}\,Institute for New Economic Thinking at the Oxford Martin School,\\ University of Oxford\\
  \textsuperscript{4}\,FLAIR, Foerster Lab for AI Research, University of Oxford\\
 \textsuperscript{5}\,Department of Engineering, University of Oxford
}

\begin{document}

\maketitle

\begin{abstract}
  Modelling how shocks propagate in supply chains is an increasingly important challenge in economics. Its relevance has been highlighted in recent years by events such as Covid-19 and the Russian invasion of Ukraine. Agent-based models (ABMs) are a promising approach for this problem. However, calibrating them is hard. We show empirically that it is possible to achieve speed ups of over 3 orders of magnitude when calibrating ABMs of supply networks by running them on GPUs and using automatic differentiation, compared to non-differentiable baselines. This opens the door to scaling ABMs to model the whole global supply network.
\end{abstract}

\section{Introduction}

Supply-chain disruptions cause macroeconomic damages \cite{IMFSC2022}, and yet, despite an established research tradition \cite{leontief1936,acemoglu2012network,baqaee_2018_cascading}, they remain difficult to model realistically \cite{CNN2025Powell}. This challenge arises from fundamental limitations in existing approaches.
Traditional economic models rely on \textit{comparative statics} and cannot capture the explicitly dynamical phenomena that characterize disruption propagation through supply networks.
Many modern machine-learning methods are similarly constrained by data scarcity: the historical record contains only a few disruptions (e.g., pandemics or natural disasters) on which to train, making it difficult to generalize across different types of shocks or network structures.

Agent-based models (ABMs)~\citep{farmer_review} offer a promising alternative to these limitations.  ABMs are a computational paradigm where a collection of agents and their interactions are modelled explicitly in a simulator. Unlike statistical methods, ABMs provide mechanistic models that recover macroscopic quantities from the bottom-up through emergent processes. ABMs are therefore a natural choice for modelling supply chains disruptions, by capturing the micro-dynamics of each firm's production and inventory management~\citep{Hallegatte2008-lq,Hallegatte2014-km,Inoue2019-ka}. 

However, ABMs are hard to calibrate. These models contain latent parameters that must be calibrated to match real-world observations according to some preferred metric. For example, firm productivity may not be directly observable and must be inferred from macroscopic quantities. Calibration requires evaluating the simulator at multiple points in parameter space --- potentially thousands or millions of times --- incurring very high computational costs when the simulator is slow to run. Furthermore, the intractable likelihood function of ABMs prevents straightforward application of Bayesian calibration methods, hindering proper uncertainty quantification.

Several approaches have emerged to mitigate these calibration challenges. These range from developing surrogate methods that mimic ABM outputs at greater speed \citep{cozzi2025learningindividualbehavioragentbased, june_hist_match}, to neural-based approaches that directly learn likelihood or posterior distributions \citep{DYER2024104827}. There has also been significant effort in tensorizing ABMs to leverage GPU acceleration, achieving substantial speedups \citep{chopra_differentiable_2023}. Furthermore, applying automatic differentiation to ABMs enables fast sensitivity analyses that can greatly accelerate calibration methods \citep{chopra_differentiable_2023, https://doi.org/10.48550/arxiv.2509.03303}.

In this work, we present a supply chain ABM implemented in \texttt{JAX} \citep{jax2018github}, a differentiable programming, high-performance numerical computing framework for Python. We demonstrate how supply chain models can be made amenable to tensorization across thousands of firm agents, achieving substantial speed-ups compared to traditional implementations. Furthermore, by leveraging \texttt{JAX}'s automatic differentiation (AD) capabilities and its probabilistic programming library \texttt{Numpyro} \citep{phan2019composable}, we apply generalized variational inference to perform efficient ABM calibration with proper uncertainty quantification over thousands of parameters.

\section{Background and Related Work}

Attempts to model production networks date back to Leontief's Input-Output framework~\cite{leontief1936}, later extended to dynamic and firm-level contexts by Long and Plosser~\cite{long_plosser_1983_rbc}, Acemo\u{g}lu \textit{et al.}~\cite{acemoglu2012network}, and Baqaee and Farhi~\cite{baqaee_2018_cascading,baqaee_farhi_2019_micro}. These models, widely used in the mainstream economic literature, study shock propagation through \textit{comparative statics}: they compute equilibrium allocations before and after an exogenous productivity shocks, and compare outcomes such as aggregate output. While useful to understand the effects of network topology on shock propagation, this approach misses dynamical features like inventory adjustment or time-dependent recovery. 

To overcome this, researchers have turned to agent-based models (ABMs) which model supply-chain dynamics from the bottom-up. Building on the ARIO model of Hallegatte~\cite{Hallegatte2008-lq,Hallegatte2014-km}, recent work has replicated real-world shock propagation such as the 2011 Tohoku earthquake~\cite{Inoue2019-ka}, the effect of the Covid-19 lockdowns in the UK's economy~\cite{Pichler2022} or the effect on the Austrian economy of an embargo on Russian gas~\cite{Pichler2024}. However, none of these works attempted a serious calibration of their models beyond parameter exploration.

Parameter estimation in ABMs has traditionally relied on Approximate Bayesian computation (ABC) methods, which sidestep intractable likelihood computation by matching summary statistics between simulated and observed data. While this approach provides a general framework for ABM calibration, it suffers from sensitivity to summary statistic selection and poor scaling in high-dimensional parameter spaces \citep{PLATT2020103859}. These limitations have motivated the development of more sophisticated methodological approaches.

One prominent strategy develops surrogate models that approximate ABM behavior while maintaining computational tractability and differentiability. By replacing the original ABM with these more amenable representations, researchers can apply established parameter estimation techniques, including Markov Chain Monte Carlo and gradient-based variational inference \cite{june_hist_match, cozzi2025learningindividualbehavioragentbased, monti_learning_2023, lamperti2017agentbasedmodelcalibrationusing}.

Neural methods represent another paradigm, directly learning the mapping from simulator parameters to posterior distributions \citep{DYER2024104827, BoeltsDeistler_sbi_2025}. These approaches offer compelling advantages through amortized inference—once trained, they provide immediate posterior estimates for new observations without retraining. However, their inability to exploit gradient information limits efficiency in high-dimensional parameter spaces.

Most recently, advances in differentiable programming have enabled a third approach that exploits the inherent computational structure of ABMs. Network-based ABMs in epidemiology have demonstrated high degrees of tensorization, achieving substantial GPU acceleration \citep{chopra_differentiable_2023}. By applying relaxation techniques to discrete randomness and control flow structures, these methods integrate ABMs directly into AD frameworks, enabling end-to-end optimization without the approximation errors of surrogate models or the gradient limitations of neural approaches \citep{https://doi.org/10.48550/arxiv.2509.03303}.

\section{Methods}
\subsection{Model}

\begin{figure}
    \centering
    \includegraphics[width=\textwidth]{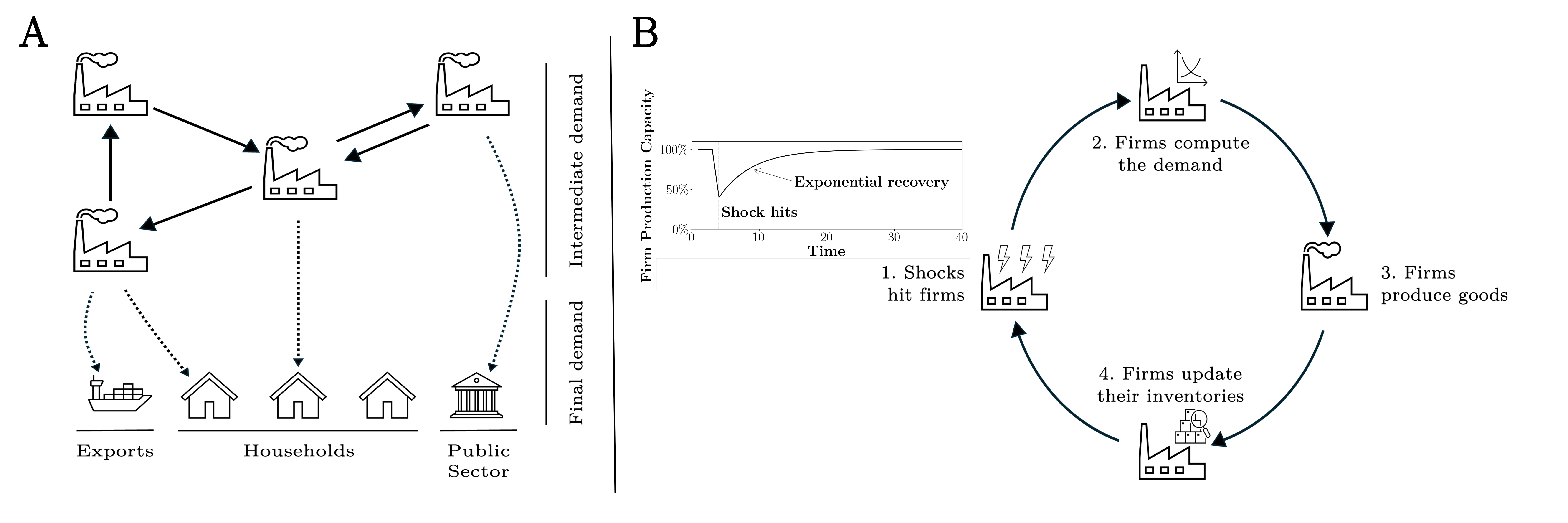}
    \caption{\textbf{A}: We model a directed production network where firms produce to serve demand from other firms, households, government, and foreign customers. \textbf{B}: In each period, an exogenous shock may limit each firm's capacity to produce, then firms receive and place orders, produce subject to input inventories and a capacity limit, and update inventories.}
    \label{fig:placeholder}
\end{figure}

We model a directed production network of $M$ firms, where each firm $i$ produces one good to serve demand from other firms (intermediate demand) as well as households, government, and foreign customers (final demand, $d_i$). Time is discrete: in each period, firms receive and place orders, produce subject to input inventories and a capacity limit, and update inventories; at the beginning of each period, an exogenous shock can decrease the maximum output that they can produce. Input requirements are encoded by a technical-coefficient matrix $\mathbf{A}$ (entries $A_{ij}$ are units of input $j$ required to produce one unit of output $i$), and inputs and outputs are tied by the formula $f\left(S_{ij}\right) = z_i\min\frac{S_{ij}}{A_{ij}}$, where $z_i$ is the productivity and set to $z_i=1$ in absence of a shock. 

Firms try to maintain a target inventory $S^{\text{target}}_{ij} = n_i S_{ij}(0)$, enough to produce for $n_i$ iterations if they don't receive supplies. Let total demand for good $i$ be $D_i(t)=d_i+\sum_k O_{k i}(t)$. Firms place orders to cover final demand and close inventory gaps, with an adjustment speed $1/\tau_i$, so
$
O_{ji}(t)=A_{ji} D_i(t)+\big[S^{\text{target}}_{ij}-S_{ji}(t-1)\big]/\tau_i.
$

The actual production is the smallest among the demand received, and the firms' production capacity, given productivity and input constraints:
$
x_i(t)=\min\left\{D_i(t-1),\; z_i(t)f\left(S_{ji}(t)\right)\right\},
$
where the productivity follows a shock-and-recovery process
$z_i(t)=\left(1-\delta_i \exp\left(-\lambda_i (t-t^*)^+\right)\right)$ for shocked firms, with $(t-t^*)^+=\max\{0,t-t^*\}$ and $t^*$  the time of the shock.
Inventories evolve by inflow minus use,
$S_{ji}(t)=S_{ji}(t-1)+Z_{ji}(t)-A_{ji}x_i(t)$,
where $Z_{ji}(t)$ denotes the realized deliveries from $j$ to $i$.
We initialize at a steady state (orders match $D_i$ and inventories sit at target) and study how shocks propagate through orders, inventories, and capacity constraints.

The model produces individual time series for each firm $x_i(t)$, but in practice, we aggregate them to obtain macroeconomic aggregates that would match regional production indices. This is obtained, e.g., by defining $y_r(t)=\sum_{i\in r} x_i(t)$ where $1\leq r\leq N$ denotes the set of firms in a given sector and a given region.

\subsection{Calibration}
Let $\mathbf{y}$ denote the aggregate observed time series of firms' outputs, $\mathbf{n} = \left[n_1, \ldots, n_N\right]$ the model parameters --- the array of firms' inventory levels --- with prior $p(\mathbf{n})$, and $\ell\!\left(\mathbf{y};\mathbf{n}\right)$ a loss function measuring the discrepancy between the observed data and the simulated results. We estimate the (generalized) posterior over parameters with Generalized Variational Inference (GVI)  \citep{knoblauch_optimization-centric_2022}, which approximates $p(\mathbf{n} \mid \mathbf{y})$ by a distribution $q^*(\mathbf{n})\in\mathcal{Q}$ solving
\begin{equation}
\label{eq:gvi}
q^*(\mathbf{n})\;=\;\arg\min_{q(\mathbf{n})\in\mathcal{Q}}\;
\mathbb{E}_{\mathbf{n}\sim q}\!\left[\ell\!\left(y;\mathbf{n}\right)\right]
\;+\; D\!\left(q(\mathbf{n})\,\Vert\,p(\mathbf{n})\right),
\end{equation}
where $D\left( \cdot \Vert p\left(\mathbf{n}\right)\right)$ penalizes deviations from the prior. In practice, we parameterize $q\left(\mathbf{n}\right)$ using a set of variational parameters $\phi$ (e.g., $\mathcal{Q}$ can be the Gaussian family \(q_\phi(\mathbf{n})=\mathcal{N}(\mathbf{n};\mu,\Sigma)\) with \(\phi=(\mu,\Sigma)\)) and minimize the objective in \eqref{eq:gvi} with stochastic gradient descent. Note that when $\ell$ is the negative log-likelihood and $D$ is the Kullback–Leibler divergence, GVI becomes the usual variational inference. For our purposes, we take $l$ to simply be the $L_2$ loss, and $D$ to be the Kullback–Leibler divergence. 
As a gradient-free baseline, we utilise a variant of Approximate Bayesian Computation (ABC), where we repeatedly sample from the prior, and keep the top $k$ samples (for us, 100) with the lowest loss under $\ell$ and $\mathbf{n}$. 
\section{Results}

\begin{figure}
  \centering
  \includegraphics[width=.9\linewidth]{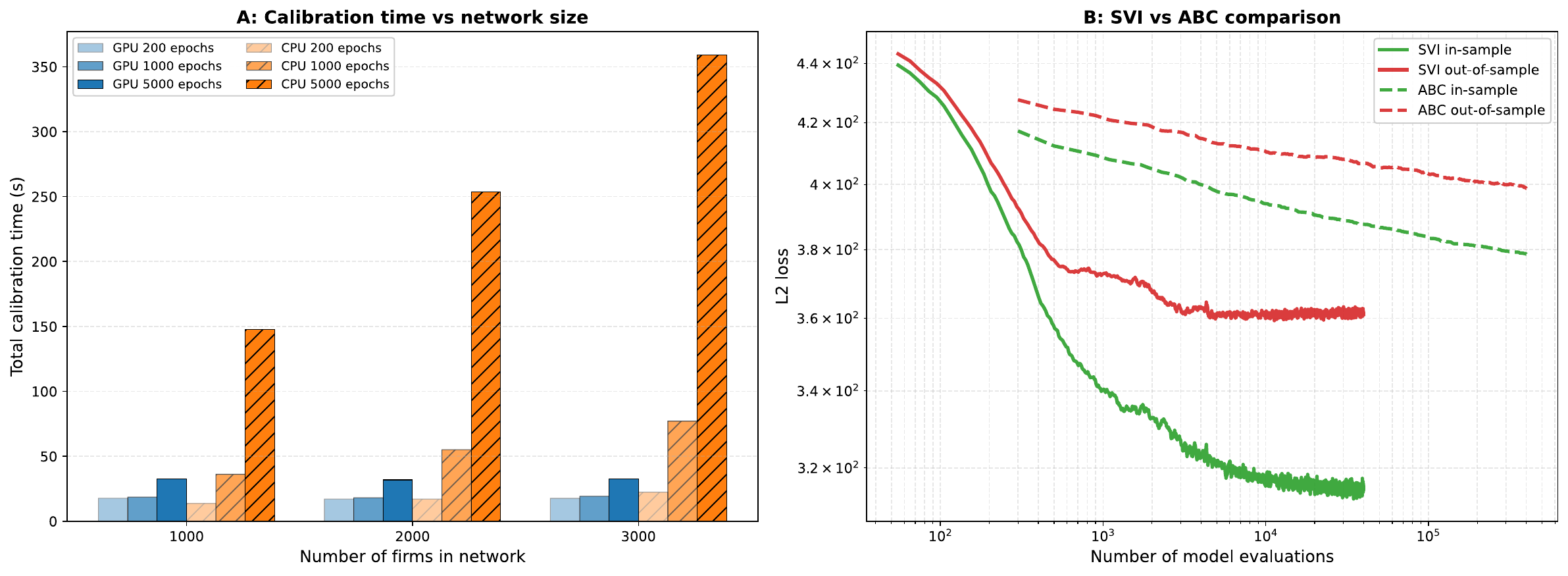}
  \caption{Two plots showing the advantages of our methodology. A: we compare the speed of SVI calibration on GPU (RTX 5090) vs CPU (Ryzen 9 9950X). B: We show how the accuracy per model evaluation decreases between SVI and ABC. }
  \label{fig:speedcomp}
\end{figure}

To demonstrate the advantages of our methodology, we run two separate experiments, shown in Figure \ref{fig:speedcomp}. Firstly, in Fig. \ref{fig:speedcomp} A, we show how our problem is naturally parallelizable by comparing the same code running on a GPU (RTX 5090) vs CPU (Ryzen 9 9950X). The benchmark used for comparison simply runs SVI for a fixed number of epochs, with different numbers of synthetic firms. The CPU-based implementation takes far longer, especially as the number of firms increase. On the other hand, the GPU-based implementation is much quicker, and doesn't increase dramatically as the number of firms increase, this is because the GPU is not using all it's parallel capacity, even at 3000 firms. As the number of firms increases, the difference between the two methods becomes only more dramatic. 

In Fig. \ref{fig:speedcomp} B, we compare to a gradient-free baseline, showing that using a gradient based method yields huge speed-ups. We do this by taking our model, which consists of 1,000 firms, and sampling some true parameters $\mathbf{n}$ from the prior, 2 for each firm. We then generate observed data from $\mathbf{n}$, and then have the two methods train on this data, utilising the true prior. To measure potential over-fitting, we compare both the ``in-sample'' loss, that is the loss to the observed data, and the ``out-of-sample'' loss, which is the average loss to observed data re-sampled from the true prior, conditioning on the true parameters $\mathbf{n}$. 

As we see from the experiment, SVI makes it possible to efficiently calibrate ABMs with 1000s of parameters, in contrast to gradient free methods like ABC. After only 300 model evaluations, SVI has a lower loss than ABC after 30,000 samples. Note that this comparison is slightly unfair, as in SVI, a model evaluation also includes a gradient pass, which adds computation that isn't measured here. However, the gradient pass takes comparable time to the forward pass, and as a result, measuring wall clock time does not meaningfully change the results. 

\section{Conclusion}
Identifying and mitigating vulnerabilities in supply chains is crucial to supporting the well-being and prosperity of the general public.
Doing this effectively requires accurate models of production networks.
Agent-based models (ABMs) are a promising approach, but are hard to calibrate. By implementing a 1,000-firm production-network ABM in \texttt{JAX}, we exploit parallelisation and automatic differentiation to perform Bayesian inference orders of magnitude faster than non-differentiable alternatives. This work not only scales ABMs but also enhances their realism—for example, by enabling simulations of price dynamics, logistics, and network restructuring in response to disruptions.

\bibliographystyle{plain}
\bibliography{biblio, arnau_citations}

@article{DYER2024104827,
title = {Black-box Bayesian inference for agent-based models},
journal = {Journal of Economic Dynamics and Control},
volume = {161},
pages = {104827},
year = {2024},
issn = {0165-1889},
doi = {https://doi.org/10.1016/j.jedc.2024.104827},
url = {https://www.sciencedirect.com/science/article/pii/S0165188924000198},
author = {Joel Dyer and Patrick Cannon and J. Doyne Farmer and Sebastian M. Schmon}
}

@article{PLATT2020103859,
title = {A comparison of economic agent-based model calibration methods},
journal = {Journal of Economic Dynamics and Control},
volume = {113},
pages = {103859},
year = {2020},
issn = {0165-1889},
doi = {https://doi.org/10.1016/j.jedc.2020.103859},
url = {https://www.sciencedirect.com/science/article/pii/S0165188920300294},
author = {Donovan Platt}
}

@article{farmer_review,
Author = {Axtell, Robert L. and Farmer, J. Doyne},
Title = {Agent-Based Modeling in Economics and Finance: Past, Present, and Future},
Journal = {Journal of Economic Literature},
Volume = {63},
Number = {1},
Year = {2025},
Month = {March},
Pages = {197–287},
DOI = {10.1257/jel.20221319},
URL = {https://www.aeaweb.org/articles?id=10.1257/jel.20221319}}

@misc{cozzi2025learningindividualbehavioragentbased,
      title={Learning Individual Behavior in Agent-Based Models with Graph Diffusion Networks}, 
      author={Francesco Cozzi and Marco Pangallo and Alan Perotti and André Panisson and Corrado Monti},
      year={2025},
      eprint={2505.21426},
      archivePrefix={arXiv},
      primaryClass={cs.AI},
      url={https://arxiv.org/abs/2505.21426}, 
}

@inproceedings{chopra_differentiable_2023,
	address = {Richland, SC},
	series = {{AAMAS} '23},
	title = {Differentiable {Agent}-based {Epidemiology}},
	isbn = {978-1-4503-9432-1},
	booktitle = {Proceedings of the 2023 {International} {Conference} on {Autonomous} {Agents} and {Multiagent} {Systems}},
	publisher = {International Foundation for Autonomous Agents and Multiagent Systems},
	author = {Chopra, Ayush and Rodríguez, Alexander and Subramanian, Jayakumar and Quera-Bofarull, Arnau and Krishnamurthy, Balaji and Prakash, B. Aditya and Raskar, Ramesh},
	year = {2023},
	note = {event-place: London, United Kingdom},
	pages = {1848--1857},
}

@article{june_hist_match,
author = {Vernon, I.  and Owen, J.  and Aylett-Bullock, J.  and Cuesta-Lazaro, C.  and Frawley, J.  and Quera-Bofarull, A.  and Sedgewick, A.  and Shi, D.  and Truong, H.  and Turner, M.  and Walker, J.  and Caulfield, T.  and Fong, K.  and Krauss, F. },
title = {Bayesian emulation and history matching of <span class="monospace">JUNE</span>},
journal = {Philosophical Transactions of the Royal Society A: Mathematical, Physical and Engineering Sciences},
volume = {380},
number = {2233},
pages = {20220039},
year = {2022},
doi = {10.1098/rsta.2022.0039},

URL = {https://royalsocietypublishing.org/doi/abs/10.1098/rsta.2022.0039},
eprint = {https://royalsocietypublishing.org/doi/pdf/10.1098/rsta.2022.0039}
}

@misc{jax2018github,
      author = {James Bradbury and Roy Frostig and Peter Hawkins and Matthew James Johnson and Chris Leary and Dougal Maclaurin and George Necula and Adam Paszke and Jake Vander{P}las and Skye Wanderman-{M}ilne and Qiao Zhang},
      title = {{JAX}: composable transformations of {P}ython+{N}um{P}y programs},
      url = {http://github.com/jax-ml/jax},
      version = {0.3.13},
      year = {2018},
}

@article{phan2019composable,
  title={Composable Effects for Flexible and Accelerated Probabilistic Programming in NumPyro},
  author={Phan, Du and Pradhan, Neeraj and Jankowiak, Martin},
  journal={arXiv preprint arXiv:1912.11554},
  year={2019}
}

@article{monti_learning_2023,
    title = {On learning agent-based models from data},
    volume = {13},
    copyright = {2023 The Author(s)},
    issn = {2045-2322},
    url = {https://www.nature.com/articles/s41598-023-35536-3},
    doi = {10.1038/s41598-023-35536-3},
    language = {en},
    number = {1},
    urldate = {2025-09-10},
    journal = {Scientific Reports},
    author = {Monti, Corrado and Pangallo, Marco and De Francisci Morales, Gianmarco and Bonchi, Francesco},
    month = jun,
    year = {2023},
    note = {Publisher: Nature Publishing Group},
    pages = {9268},
}

@misc{lamperti2017agentbasedmodelcalibrationusing,
      title={Agent-Based Model Calibration using Machine Learning Surrogates}, 
      author={Francesco Lamperti and Andrea Roventini and Amir Sani},
      year={2017},
      eprint={1703.10639},
      archivePrefix={arXiv},
      primaryClass={q-fin.EC},
      url={https://arxiv.org/abs/1703.10639}, 
}

@article{BoeltsDeistler_sbi_2025,
  doi = {10.21105/joss.07754},
  url = {https://doi.org/10.21105/joss.07754},
  year = {2025},
  publisher = {The Open Journal},
  volume = {10},
  number = {108},
  pages = {7754},
  author = {Jan Boelts and Michael Deistler and Manuel Gloeckler and Álvaro Tejero-Cantero and Jan-Matthis Lueckmann and Guy Moss and Peter Steinbach and Thomas Moreau and Fabio Muratore and Julia Linhart and Conor Durkan and Julius Vetter and Benjamin Kurt Miller and Maternus Herold and Abolfazl Ziaeemehr and Matthijs Pals and Theo Gruner and Sebastian Bischoff and Nastya Krouglova and Richard Gao and Janne K. Lappalainen and Bálint Mucsányi and Felix Pei and Auguste Schulz and Zinovia Stefanidi and Pedro Rodrigues and Cornelius Schröder and Faried Abu Zaid and Jonas Beck and Jaivardhan Kapoor and David S. Greenberg and Pedro J. Gonçalves and Jakob H. Macke},
  title = {sbi reloaded: a toolkit for simulation-based inference workflows},
  journal = {Journal of Open Source Software}
}

@article{leontief1936,
  title={Quantitative Input and Output Relations in the Economic System of the United States},
  author={Leontief, Wassily},
  journal={The Review of Economics and Statistics},
  volume={18},
  number={3},
  pages={105--125},
  year={1936}
}

@article{acemoglu2012network,
  title        = {The Network Origins of Aggregate Fluctuations},
  author       = {Acemoglu, Daron and Carvalho, Vasco M. and Ozdaglar, Asuman and Tahbaz-Salehi, Alireza},
  journal      = {Econometrica},
  volume       = {80},
  number       = {5},
  pages        = {1977--2016},
  year         = {2012},
  doi          = {10.3982/ECTA9623}
}

@article{baqaee_farhi_2019_micro,
  title        = {The Macroeconomic Impact of Microeconomic Shocks: Beyond Hulten's Theorem},
  author       = {Baqaee, David Rezza and Farhi, Emmanuel},
  journal      = {Econometrica},
  volume       = {87},
  number       = {4},
  pages        = {1155--1203},
  year         = {2019},
  doi          = {10.3982/ECTA15202}
}

@article{baqaee_2018_cascading,
  title        = {Cascading Failures in Production Networks},
  author       = {Baqaee, David Rezza},
  journal      = {Econometrica},
  volume       = {86},
  number       = {5},
  pages        = {1819--1838},
  year         = {2018},
  doi          = {10.3982/ECTA15280}
}

@article{Pichler2024,
  title = {Economic impacts of a drastic gas supply shock and short-term mitigation strategies},
  volume = {227},
  ISSN = {0167-2681},
  url = {http://dx.doi.org/10.1016/j.jebo.2024.106750},
  DOI = {10.1016/j.jebo.2024.106750},
  journal = {Journal of Economic Behavior \& Organization},
  publisher = {Elsevier BV},
  author = {Pichler,  Anton and Hurt,  Jan and Reisch,  Tobias and Stangl,  Johannes and Thurner,  Stefan},
  year = {2024},
  month = nov,
  pages = {106750}
}

@ARTICLE{Hallegatte2008-lq,
  title     = "An adaptive regional input-output model and its application to
               the assessment of the economic cost of Katrina",
  author    = "Hallegatte, St{\'e}phane",
  journal   = "Risk Anal.",
  publisher = "Wiley",
  volume    =  28,
  number    =  3,
  pages     = "779--799",
  month     =  jun,
  year      =  2008,
  copyright = "http://onlinelibrary.wiley.com/termsAndConditions\#vor",
  language  = "en"
}

@ARTICLE{Hallegatte2014-km,
  title     = "Modeling the role of inventories and heterogeneity in the
               assessment of the economic costs of natural disasters",
  author    = "Hallegatte, St{\'e}phane",
  journal   = "Risk Anal.",
  publisher = "Wiley",
  volume    =  34,
  number    =  1,
  pages     = "152--167",
  month     =  jan,
  year      =  2014,
  copyright = "http://onlinelibrary.wiley.com/termsAndConditions\#vor",
  language  = "en"
}

@ARTICLE{Inoue2019-ka,
  title     = "Firm-level propagation of shocks through supply-chain networks",
  author    = "Inoue, Hiroyasu and Todo, Yasuyuki",
  journal   = "Nat. Sustain.",
  publisher = "Springer Science and Business Media LLC",
  volume    =  2,
  number    =  9,
  pages     = "841--847",
  month     =  aug,
  year      =  2019,
  copyright = "https://www.springernature.com/gp/researchers/text-and-data-mining",
  language  = "en"
}

@misc{https://doi.org/10.48550/arxiv.2509.03303,
  doi = {10.48550/ARXIV.2509.03303},
  url = {https://arxiv.org/abs/2509.03303},
  author = {Quera-Bofarull,  Arnau and Bishop,  Nicholas and Dyer,  Joel and Ornia,  Daniel Jarne and Calinescu,  Anisoara and Farmer,  Doyne and Wooldridge,  Michael},
  title = {Automatic Differentiation of Agent-Based Models},
  publisher = {arXiv},
  year = {2025},
  copyright = {arXiv.org perpetual,  non-exclusive license}
}

@article{Pichler2022,
  title = {Forecasting the propagation of pandemic shocks with a dynamic input-output model},
  volume = {144},
  ISSN = {0165-1889},
  url = {http://dx.doi.org/10.1016/j.jedc.2022.104527},
  DOI = {10.1016/j.jedc.2022.104527},
  journal = {Journal of Economic Dynamics and Control},
  publisher = {Elsevier BV},
  author = {Pichler,  Anton and Pangallo,  Marco and del Rio-Chanona,  R. Maria and Lafond,  Fran\c{c}ois and Farmer,  J. Doyne},
  year = {2022},
  month = nov,
  pages = {104527}
}

@misc{CNN2025Powell,
  author       = {{CNN Business}},
  title        = {Federal Reserve Chair Jerome Powell: "We don't have the kind of tools to deal with supply chain problems"},
  howpublished = {\url{http://archive.today/2025.05.07-212523/https://edition.cnn.com/business/live-news/federal-reserve-interest-rate-05-07-24}},
  year         = {2025},
  note         = {Live transcript, May 7, 2025. Accessed: 2025-10-01},
  institution  = {CNN Business},
}

@article{long_plosser_1983_rbc,
  title        = {Real Business Cycles},
  author       = {Long, John B., Jr. and Plosser, Charles I.},
  journal      = {Journal of Political Economy},
  volume       = {91},
  number       = {1},
  pages        = {39--69},
  year         = {1983},
  doi          = {10.1086/261128}
}

@article{IMFSC2022,
	author = {Celasun, Oya and Mineshima, Aiko and Hansen, Niels-Jakob and Zhou, Jing and Spector, Mariano},
	journal = {IMF Working Papers},
	doi = {10.5089/9798400202476.001},
	issn = {1018-5941},
	number = {031},
	year = {2022},
	month = {2},
	pages = {1},
	publisher = {International Monetary Fund (IMF)},
	title = {Supply {Bottlenecks}: Where, {Why}, {How} {Much}, and {What} {Next}?},
	url = {http://dx.doi.org/10.5089/9798400202476.001},
	volume = {2022},
}

@article{knoblauch_optimization-centric_2022,
    title = {An {Optimization}-centric {View} on {Bayes}' {Rule}: {Reviewing} and {Generalizing} {Variational} {Inference}},
    volume = {23},
    url = {http://jmlr.org/papers/v23/19-1047.html},
    number = {132},
    journal = {Journal of Machine Learning Research},
    author = {Knoblauch, Jeremias and Jewson, Jack and Damoulas, Theodoros},
    year = {2022},
    pages = {1--109},
}

\end{document}